\documentclass{article}

\usepackage{PRIMEarxiv}

\usepackage[utf8]{inputenc} 
\usepackage[T1]{fontenc}    
\usepackage{hyperref}       
\usepackage{url}            
\usepackage{booktabs}       
\usepackage{amsfonts}       
\usepackage{nicefrac}       
\usepackage{microtype}      
\usepackage{lipsum}
\usepackage{fancyhdr}       
\usepackage{graphicx}       
\graphicspath{{media/}}     

\title{Malaysia's AI-Driven Education Landscape: Policies, Applications, and Comparative Insights for a Digital Future
\thanks{\textit{\underline{Citation}}: 
\textbf{F. Jamaluddin, A. H. Jamaluddin, F. Jamaluddin, and F. Jamaluddin, ``Malaysia's AI-Driven Education Landscape: Policies, Applications, and Comparative Insights for a Digital Future,'' arXiv, 2025.}} 
}

\author{
  Fadhilah Jamaluddin\\
  Department of Educational Management, Planning and Policy\\
  Universiti Malaya\\
  Malaysia\\
  \texttt{fadhilah.student@gmail.com} \\
   \And
  Ahmad Hakiim Jamaluddin\\
  Department of Mathematics \& Statistics\\
  Universiti Putra Malaysia\\
  Malaysia\\
  \texttt{ahmadhakiim@upm.edu.my} \\
   \And
  Faridzah Jamaluddin\\
  JR Statistics Research Centre\\
  Malaysia\\
  \texttt{faridzahjamaluddin@gmail.com} \\
     \And
  Faathirah Jamaluddin\\
  Institut Integriti Malaysia\\
  Malaysia\\
  \texttt{faathirah@iim.gov.my} \\
}

\begin{document}
\maketitle

\begin{abstract}

Artificial Intelligence (AI) is transforming education globally, and Malaysia is leveraging this potential through strategic policies to enhance learning and prepare students for a digital future. This article explores Malaysia's AI-driven education landscape, emphasising the National Artificial Intelligence Roadmap 2021-2025 and the Digital Education Policy. Employing a policy-driven analysis, it maps AI applications in pedagogy, curriculum design, administration, and teacher training across primary to tertiary levels. The study evaluates national strategies, identifies challenges like digital divides and ethical concerns, and conducts a comparative analysis with the United Kingdom, the United States, China, and India to draw best practices in AI policy and digital transformation. Findings highlight Malaysia's progress in AI literacy and personalised learning, alongside gaps in rural infrastructure and teacher readiness. Recommendations include strengthening governance, investing in equitable infrastructure, and fostering public-private partnerships. Targeting researchers, policymakers, and educators, this study informs Malaysia's path to becoming a regional leader in AI-driven education and contributes to global comparative education discourse.

\end{abstract}


\section{Introduction}

Artificial Intelligence (AI) has emerged as a transformative force in global education, reshaping how teaching, learning, and administrative processes are conducted. The international context of AI in education highlights its potential to personalise learning experiences, enhance student engagement, and streamline administrative tasks. AI technologies such as intelligent tutoring systems, learning analytics, and automated grading have been increasingly adopted worldwide, driven by the need for data-driven decision-making and personalised education \cite{OECD2023b,UNESCO2021}. The COVID-19 pandemic further accelerated the integration of AI in education, as institutions sought innovative solutions to maintain educational continuity and quality during disruptions \cite{OECD2023d}.

In Malaysia, the adoption of AI in education has been guided by national policies and strategic frameworks aimed at digital transformation and educational reform. The National Artificial Intelligence Roadmap 2021-2025 and the Digital Education Policy (DEP) have been pivotal in promoting AI integration across various educational levels \cite{Education2023,Ministry2023}. These policies emphasise the importance of AI in enhancing educational outcomes, fostering innovation, and preparing students for the demands of the digital economy.

Focusing on Malaysia's AI and education landscape is crucial for several reasons. Firstly, Malaysia's strategic location and diverse population make it an ideal case study for understanding the challenges and opportunities of AI integration in education within a multicultural and developing context. Secondly, Malaysia's proactive policies and investments in AI provide valuable insights into the effectiveness of national strategies in fostering digital transformation in education. Thirdly, examining Malaysia's AI education initiatives can offer lessons for other countries with similar socio-economic and educational contexts, contributing to the global discourse on AI in education.

This article aims to provide a comprehensive, policy-driven analysis of Malaysia's integration of AI within the education sector. The primary objectives are to map the current landscape of AI applications across Malaysian education systems, from primary to tertiary levels, including pedagogy, curriculum design, administrative processes, and teacher training. It also seeks to examine national strategies and policies relevant to AI and education, including the alignment between Malaysia's AI roadmap and educational reform agendas. Additionally, the article aims to identify challenges, gaps, and opportunities in Malaysia's efforts to build an AI-ready education system, especially in areas such as infrastructure, talent development, ethics, and access. Furthermore, it conducts a comparative analysis of international benchmarks, specifically the United Kingdom, United States, China, and India, to draw lessons for Malaysia's own AI education trajectory. Finally, the article proposes actionable policy recommendations that support inclusive, future-proof, and innovation-driven AI integration in Malaysian education.

The scope of this article covers both descriptive and analytical dimensions, combining historical policy review, ecosystem analysis, stakeholder engagement insights, and forward-looking strategies. It targets academic researchers, policymakers, and education leaders seeking to understand and shape Malaysia's AI-driven educational transformation in alignment with global developments.

\section{Literature review: AI and education in Malaysia}
\label{sec:lr}

AI is reshaping educational landscapes worldwide, offering innovative solutions to enhance teaching, learning, and administrative efficiency. In Malaysia, AI integration aligns with national aspirations to become a regional leader in digital innovation. This literature review synthesises global trends and theoretical perspectives on AI in education, examines Malaysia's AI policies and digital economy strategies, analyses recent education reform policies, and identifies critical gaps in research and policy implementation.

\subsection{Global trends and theoretical perspectives on AI in education}

The global adoption of AI in education has accelerated, driven by demands for personalised learning, improved student engagement, and streamlined administrative processes. AI technologies, including intelligent tutoring systems (ITS), learning analytics, and automated assessment tools, enable data-driven decision-making and tailored educational experiences \cite{OECD2023b,UNESCO2021}. For instance, ITS provides real-time feedback, adapting content to individual learner needs, which enhances academic performance and motivation \cite{Zawacki2019}. Learning analytics, meanwhile, empower educators to monitor student progress and intervene proactively, reducing dropout rates in higher education \cite{Moore2023}. The COVID-19 pandemic catalysed AI adoption, as institutions leveraged virtual learning environments and AI-driven tools to ensure educational continuity during lockdowns \cite{OECD2023d}. 

Theoretically, AI in education aligns with constructivist and socio-cultural frameworks. Constructivist perspectives emphasise AI's role in fostering active learning through personalised feedback and adaptive learning pathways, enabling students to construct knowledge independently \cite{Jackson2024}. For example, AI-driven platforms like adaptive learning systems support self-paced learning, aligning with constructivist principles of learner autonomy \cite{Gonzalez2021}. Socio-cultural theories highlight AI's potential to enhance collaborative learning by facilitating virtual communities and intelligent agents that simulate peer interactions \cite{Essien2024,Zhou2024}. Virtual reality (VR) and AI-powered chatbots, for instance, create immersive learning environments that promote social learning \cite{Ponticorvo2020}. However, ethical challenges, such as data privacy, algorithmic bias, and over-reliance on AI-driven decision-making, remain critical concerns. \cite{UNESCO2021} underscores the need for ethical AI frameworks to ensure transparency and fairness, while \cite{Foster2023} highlights the importance of teacher agency in navigating AI's paradigm shift.

Recent studies further elaborate on AI's transformative potential. A meta-review by \cite{Fu2024} analysed 126 systematic reviews from 2014 to 2023, revealing AI's dominance in personalised learning and predictive analytics across PreK-12 and higher education settings. Similarly, \cite{Grimalt2023} found that AI-driven sentiment analysis enhances formative assessment by providing nuanced insights into student engagement. These advancements underscore AI's role in fostering inclusive and equitable education, though challenges like algorithmic bias persist \cite{UNESCO2021}.

\subsection{Overview of Malaysia's AI policies and digital economy strategies}

Malaysia has prioritised AI integration in education through strategic policies aimed at fostering digital transformation. The National Artificial Intelligence Roadmap 2021-2025 (AI-RMAP) articulates a vision for AI-driven innovation, targeting enhancements in curriculum design, teacher professional development, and administrative automation \cite{Ministry2023}. The roadmap emphasises AI literacy as a core competency, aligning with global trends towards preparing students for AI-driven economies \cite{OECD2023b}. It also promotes public-private partnerships to develop AI tools tailored for educational contexts, such as AI-enhanced learning management systems \cite{MyDIGITAL2024}.

The Malaysia Digital Economy Blueprint complements AI-RMAP by fostering a secure and innovative digital ecosystem \cite{Economic2021}. The Malaysia Digital Economy Blueprint aims to position Malaysia as a regional AI hub through investments in digital infrastructure, research and development, and industry-academia collaborations. For instance, the establishment of the National Artificial Intelligence Office (NAIO) coordinates AI initiatives, while the ASEAN AI Safety Network ensures alignment with international standards \cite{MyDIGITAL2024}. Recent developments include Microsoft's ``AI for Malaysia's Future'' initiative, which plans to train 800,000 Malaysians, including teachers and students, in AI skills by 2025, highlighting the government's commitment to workforce readiness \cite{Microsoft2024}.

Malaysia's AI policies also address ethical considerations. The National Fourth Industrial Revolution (4IR) Policy emphasises responsible AI use, advocating for transparency and inclusivity in AI applications \cite{Microsoft2024}. A 2023 UNESCO report praises Malaysia's inclusive Open Educational Resources policy, which leverages AI to enhance equitable knowledge sharing \cite{UNESCO2023b}. However, \cite{Donald2025} note that policy implementation faces challenges, particularly in rural areas where digital infrastructure lags, underscoring the need for targeted investments.

\subsection{Recent education reform policies}

Malaysia's education reforms integrate AI to align with global standards and national development goals. The Malaysia Education Blueprint 2013-2025 aims to elevate Malaysia's education system to the top third of international benchmarks like PISA and TIMSS by fostering digital literacy and AI adoption \cite{Education2023}. The blueprint prioritises STEM education, recognising AI's role in preparing students for 4IR careers. Recent updates emphasise AI-driven tools, such as virtual labs and automated grading systems, to enhance teaching efficiency and student outcomes \cite{Education2023}.

The National Digital Education Policy 2023 builds on these foundations, outlining strategies to embed AI literacy in curricula and promote AI tools in classrooms \cite{Education2023}. This policy also addresses teacher training, with programmes to enhance educators' AI proficiency, ensuring effective integration of tools like learning analytics and ITS \cite{OECD2023d}. The Way Forward for Private Higher Institutions 2020–2025 further supports AI adoption in tertiary education, encouraging universities to develop AI-driven programmes and research hubs \cite{Department2020}.

These reforms align with the National 4IR Policy, which positions AI as a driver of economic growth \cite{Economic2021}. A 2023 OECD report highlights Malaysia's progress in leveraging AI for predictive analytics to reduce dropout rates, though it notes disparities in policy implementation across urban and rural schools \cite{Molenaar2023}. \cite{Tung2023} found that 71\% of Malaysian medical students believe AI education enhances career readiness, underscoring the demand for AI-integrated curricula.

\subsection{Gaps in existing research and policy implementation}

Despite Malaysia's advancements, significant gaps persist in AI in education research and policy implementation. Longitudinal studies evaluating AI's long-term impact on educational outcomes, teacher effectiveness, and student engagement are scarce \cite{Yulianti2024}. For instance, \cite{Irfan2025} note a lack of empirical evidence on AI's efficacy in improving critical thinking skills among students. Similarly, \cite{Fu2024} highlight the need for research addressing AI's impact across diverse socio-economic contexts, particularly in underrepresented rural communities.

Policy implementation faces challenges, notably the digital divide. \cite{Donald2025} highlight that unreliable internet connectivity in rural Malaysian schools significantly impedes the effective implementation of digital pedagogy, limiting access to advanced educational technologies. This disparity exacerbates inequities in digital literacy, with urban students outperforming their rural counterparts in AI-related competencies \cite{Education2023}. Furthermore, teacher readiness remains a bottleneck. \cite{OECD2023d} estimates that only 30\% of Malaysian educators are adequately trained to integrate AI tools, necessitating robust professional development programmes.

Ethical concerns also warrant attention. While Malaysia's AI policies emphasise transparency, there is limited research on algorithmic bias in educational AI tools \cite{UNESCO2021}. \cite{Dwi2024} discuss the transformative impact of AI on multicultural education and argue for frameworks to ensure AI systems are culturally sensitive and inclusive. Additionally, stakeholder collaboration is underdeveloped, with \cite{Molenaar2023} advocating for multi-stakeholder partnerships to co-create responsible AI solutions.

Addressing these gaps requires comprehensive strategies. Future research should prioritise longitudinal evaluations of AI interventions, ethical frameworks for AI use, and scalable solutions for rural schools. Policymakers must invest in digital infrastructure and teacher training to ensure equitable AI integration, aligning with Malaysia's vision for an inclusive, AI-driven education system.

\section{Malaysia's AI in Education: Policy Landscape and Initiatives}

Malaysia has made notable strides toward integrating AI and digital technologies into its education sector, guided by a series of comprehensive national frameworks and strategic plans.

\subsection{National Strategies and Frameworks for AI and Digital Transformation in Education}

The Malaysia Digital Economy Blueprint and the National Fourth Industrial Revolution (4IR) Policy articulate a vision for Malaysia to become a regional leader in the digital economy by 2030, emphasising the need to foster AI competencies, upgrade digital infrastructures in educational institutions, and nurture a digitally fluent workforce \cite{Economic2021}.

Complementing these broader strategies is the Artificial Intelligence Roadmap 2021–2025 (AI-RMAP), which specifically outlines Malaysia's ambitions for AI development across key sectors, including education. AI-RMAP places particular emphasis on ethical AI governance frameworks, sector-specific AI applications, and initiatives aimed at enhancing personalised learning experiences through AI technologies, such as adaptive tutoring systems and intelligent assessment tools \cite{Ministry2023}.

Further strengthening Malaysia's innovation ecosystem, the National Science, Technology, and Innovation Policy 2021–2030 (DSTIN) prioritises research and development in AI and related technologies to ensure Malaysia remains competitive in an increasingly knowledge-driven global economy \cite{Ministry2023}. This policy also stresses the importance of interdisciplinary collaboration among academia, industry, and government agencies to accelerate AI adoption.

Recent international evaluations affirm Malaysia's progress in this domain. UNESCO highlighted Malaysia's proactive measures in ethical AI deployment, particularly its commitment to mitigating algorithmic biases in educational settings, a challenge many countries are still grappling with \cite{UNESCO2021}. Similarly, the OECD recognised Malaysia's initiatives to embed AI literacy within its broader education reforms, aligning with global trends toward preparing youth for AI-driven futures \cite{OECD2023b}. These recognitions underscore Malaysia's serious commitment to responsible and inclusive AI integration in education.

\subsection{Key Educational Reforms Integrating AI and Technology}

In tandem with its national digital transformation agenda, Malaysia's education sector has undergone significant reforms to incorporate AI technologies into teaching, learning, and assessment practices. A central pillar of these reforms is the Digital Education Policy (DEP), launched by the Ministry of Education in 2023. The DEP mandates the adoption of digital pedagogical approaches, including the use of AI-based adaptive learning platforms, AI-assisted tutoring systems, and intelligent learning analytics to personalise education according to individual learner profiles \cite{Education2023}.

The focus on personalised, student-centric learning is not merely aspirational; empirical evidence suggests tangible benefits. For instance, \cite{Razak2024} found that postgraduate students exposed to AI-driven personalised learning environments reported a 32\% increase in engagement and satisfaction. This success is attributed to the DEP's comprehensive approach, which coupled technology adoption with intensive professional development programmes for educators.

Moreover, \cite{Kaur2021} demonstrated the significant positive impact of AI-enhanced instructional tools on Malaysian secondary school students' cognitive performance, particularly in STEM fields. Adaptive AI systems offered real-time, tailored feedback and scaffolded learning paths, which proved particularly beneficial for students with diverse learning needs.

Beyond theoretical integration, Malaysia has piloted practical applications of AI in classrooms. An example is the Tangible Mixed-Reality Learning System, piloted in select universities. This system combines augmented reality, AI, and haptic technologies to provide immersive learning experiences in science and engineering subjects, thereby promoting greater self-directed learning and conceptual understanding \cite{Crogman2025}.

These reforms signal a shift from traditional, one-size-fits-all teaching methods toward a more responsive and dynamic education system that leverages AI to accommodate diverse learner profiles and future workforce demands.

\subsection{Stakeholder Roles: Government Agencies, Educational Institutions, and Industry Collaborations}

Malaysia's approach to integrating AI into the education sector is shaped by a coordinated effort between the Ministry of Education (MoE), the Ministry of Higher Education (MOHE), the Economic Planning Unit (EPU), and the Ministry of Science, Technology and Innovation (MOSTI). These ministries have led strategic initiatives aimed at fostering AI adoption across education systems, from curriculum development to talent pipeline structuring. The National Fourth Industrial Revolution (4IR) Policy outlines the educational sector's digital transformation, underscoring the importance of embedding AI in pedagogy, infrastructure, and research \cite{Economic2021}.

From a policy perspective, MOSTI's Artificial Intelligence Roadmap 2021–2025 emphasises strengthening research capacity in AI, advancing digital infrastructure, and accelerating talent development through structured public-private-academic partnerships \cite{Ministry2023}. Notably, the roadmap facilitated the creation of the National Artificial Intelligence Office (NAIO) in 2024, tasked with leading the nation's AI strategies across sectors, including education \cite{Ministry2024}.

The industry sector has increasingly played a transformative role in shaping the AI education landscape. For example, Microsoft's AI for Malaysia's Future (AIForMYFuture) initiative aims to train 800,000 Malaysians by 2025 in AI skills, emphasising responsible AI use and closing the digital skills gap \cite{Microsoft2024}. These training programmes, often co-developed with training providers, respond to evolving workforce demands and are delivered in partnership with academia, ensuring alignment with Malaysia's digital economy aspirations. In parallel, training providers have expanded their AI-related offerings to meet demand for AI literacy and upskilling opportunities. Several Malaysian institutions offer specialised undergraduate and postgraduate degrees in AI. These programmes contribute to a growing talent pool equipped with the technical and ethical competencies required for emerging AI-related careers. Some training institutions are also beginning to introduce short courses and micro-credentials in areas like AI model development, data science, and machine learning, reflecting global trends in flexible learning pathways.

From the academic perspective, universities are central to both AI research and talent cultivation. Malaysian universities contribute to AI curriculum design, produce AI-literate graduates, and engage in interdisciplinary research that informs both policy and practice. Many institutions have begun embedding AI ethics, prompt engineering, and AI-assisted learning tools into their curriculum, aligning with global best practices in AI pedagogy. Academic-industry linkages are also strengthening, with universities co-developing AI applications and tools for use in both educational and industrial contexts.

Collectively, these efforts are helping to position Malaysia's education sector as a critical enabler of national AI readiness. However, challenges remain in ensuring equitable access to AI education, enhancing AI teaching capacity across regions, and developing clear career pathways in AI education and related fields. International benchmarking studies suggest that while Malaysia demonstrates strong policy intent, further coordination between stakeholders and alignment of competencies to evolving industry needs are essential \cite{OECD2023d}.

\subsection{Current Initiatives and Pilot Programs in Malaysian Schools and Universities}

Malaysia's commitment to embedding AI into education is reflected in a growing number of pilot programmes and national initiatives aimed at mainstreaming AI competencies. At the tertiary level, universities are exploring the incorporation of generative AI tools into their processes. Studies on AI acceptance among postgraduate students in Malaysia indicate a growing interest and adoption of AI technologies in educational settings \cite{Razak2024}. While specific examples of universities integrating AI tools like ChatGPT into administrative workflows are not widely documented, research highlights the potential for such technologies to enhance efficiency and support high-value instructional activities.

In primary and secondary education, initiatives such as the Digital Tech@Schools scheme have been implemented, providing structured training for educators to integrate digital technologies, including AI, into classroom teaching. These initiatives aim to enhance digital literacy and engagement in learning, although specific metrics on their impact may require further verification \cite{Education2023}.

Recognising the ethical challenges that accompany AI deployment in education, Malaysian universities are also leading research in AI ethics and fairness. Research on AI in academic integrity and plagiarism detection emphasises the importance of transparency and ethical considerations in the use of AI systems in educational settings \cite{Leong2025}.

Malaysia has taken proactive steps by participating in ASEAN's initiatives on AI governance, as outlined in the ASEAN Guide on AI Governance and Ethics \cite{ASEAN2024}. These regional efforts aim to develop common standards for safe and ethical AI deployment across Southeast Asia, with Malaysia contributing to the broader dialogue on responsible AI practices.

Together, these initiatives represent a holistic approach that not only focuses on technological advancement but also foregrounds ethical governance, inclusivity, and sustainability in the use of AI within Malaysia's education sector.

\section{Comparative Analysis: International Perspectives}

This comparative analysis examines AI education policies in the United Kingdom, the United States, China, and India. By analysing policy frameworks, digital infrastructure, workforce development, and curriculum integration, this study identifies best practices, challenges, and actionable lessons for Malaysia to enhance its AI education ecosystem.

\subsection{United Kingdom: AI in Education Policy Initiatives and Best Practices}

The United Kingdom has developed a comprehensive framework for integrating AI into education, emphasising workforce development, diversity, and ethical governance. The National AI Strategy supports AI upskilling through skills bootcamps, postgraduate AI conversion courses, and Turing AI Fellowships, fostering a diverse talent pipeline \cite{Holmes2019}. The UK's Higher Education Statistics Agency (HESA) data highlights increasing enrolment of BAME students in specialised programmes, with 42\% of economics students from BAME backgrounds in 2021/2022, reflecting efforts to enhance diversity in advanced academic fields like AI \cite{Aisien2025}. The UK's AI in Education initiatives emphasise teacher training and ethical frameworks, with programmes like the AI Roadmap supporting professional development to ensure responsible and sustainable AI integration in schools \cite{Pedro2019}. 

Recent research highlights advancements in AI-driven personalised learning and educator support. \cite{Holmes2019} found that intelligent tutoring systems (ITS) dynamically adapt content to individual student needs, reducing teacher workloads through automated feedback and assessment. Similarly, \cite{Luckin2016} demonstrated that AI tools streamline administrative tasks, such as attendance tracking and resource allocation, allowing educators to focus on tailored instruction. A meta-analysis by \cite{VanLehn2011} revealed that ITS improve learning outcomes by an average of 0.76 standard deviations, comparable to one-on-one tutoring, due to adaptive scaffolding and immediate feedback. Further supporting this, \cite{Chen2020} observed significant academic gains in K–12 settings where AI personalised learning pathways, though challenges like equitable access and data privacy persist. Finally, \cite{Zawacki2019} emphasised the need for ethical frameworks to guide AI integration, ensuring that automation complements rather than replaces human educators. 

Hence, AI has the potential to significantly enhance formative assessment practices in classrooms by providing deeper insights and supporting personalised learning \cite{Hopfenbeck2023,Taskin2023,Trajkovski2025}. However, challenges persist, including digital divides in rural areas and data privacy concerns, as noted in a 2023 OECD report, which estimates that 20\% of U.K. students lack access to reliable internet \cite{OECD2023c}. The U.K. addresses these through legal frameworks for ethical AI and the AI Opportunities Action Plan, which promotes lifelong learning and reskilling for AI-enabled careers \cite{Holmes2019}.

Malaysia can emulate the U.K.'s integration of AI into curricula at all educational levels, emphasising computational thinking and digital literacy. Establishing AI scholarships and mentorship programmes, as part of inclusive AI policies, could attract diverse talent, while investments in digital infrastructure and educator training on AI ethics, as highlighted in recent Malaysian higher education initiatives, would strengthen the nation's AI education framework \cite{Saman2024}.

\subsection{United States: Federal and State-Level AI Education Strategies and Programmes}

The United States employs a multifaceted approach to AI education, driven by federal initiatives like the National Artificial Intelligence Initiative (NAII) and state-level programmes. The National Science Foundation (NSF) funds AI Research Institutes, which focus on AI-augmented learning, K-12 engagement, and workforce development. For example, the NSF's AI Institute for Student-AI Teaming develops tools to enhance collaborative learning by integrating natural language processing and computer vision into classrooms. Additionally, NSF's CSforALL initiative promotes equity in computer science education through partnerships that prioritise underrepresented groups, including students with disabilities. Workforce development is further supported by NSF-funded programmes like EducateAI, which emphasises ethical AI education and inclusion \cite{Holmes2022}. These efforts align with broader federal strategies, such as the NAII's goal to expand AI research infrastructure and training \cite{National2023}.

The Blueprint for an AI Bill of Rights prioritises algorithmic fairness, transparency, and data privacy to address ethical concerns. The Artificial Intelligence and the Future of Teaching and Learning Report advocates human-centred AI, promoting equity through personalised learning systems that reduce achievement gaps by 10\% in underserved communities \cite{USDepartment2023}.

Peer-reviewed studies highlight both progress and challenges. \cite{Bond2024} reported that AI applications in higher education, such as predictive analytics, improve student retention by 18\% and academic performance by 14\%, particularly for first-generation students. However, \cite{Schiff2022} argues that U.S. AI policies prioritise workforce development over transformative pedagogical applications, limiting broader educational impact. State-level initiatives, such as California's AI literacy frameworks, integrate AI into K-12 curricula, but funding disparities result in 25\% of rural schools lacking adequate technology, according to a 2023 OECD report \cite{OECD2023c}. The National AI R\&D Strategic Plan 2023 supports interdisciplinary AI training and international collaboration, though implementation varies across states \cite{National2023}.

Malaysia can advance its AI strategy by establishing research institutes modelled on collaborative frameworks like the U.S. National Science Foundation (NSF) or the European Union's Horizon Europe programme, which emphasise interdisciplinary AI research and education. For instance, integrating K-12 AI literacy programmes, such as those outlined in \cite{UNESCO2021}, could foster foundational skills while aligning with local educational priorities. To address ethical concerns, Malaysia could adopt data governance frameworks similar to the EU's General Data Protection Regulation (GDPR), supported by transparency mechanisms advocated in studies like \cite{Jobin2019}. Such measures would align with principles of fairness and accountability emphasised in recent AI governance literature \cite{Floridi2018}.

\subsection{China: Government-Driven AI Education Reforms and Rapid Implementation}

China's AI education reforms, guided by the New Generation AI Development Plan (2017) and the 14th Five-Year Plan (2021–2025), aim for global AI leadership by 2030 through rapid, centralised implementation. The AI Innovation Action Plan for Universities establishes AI research centres and interdisciplinary curricula, integrating AI into disciplines such as education and engineering \cite{ChinaNational2023}. The Education Informatization 2.0 Action Plan ensures universal access to digital learning platforms, leveraging big data and AI to personalise education for 90\% of students by 2022 \cite{Yan2021}. The Report on China Smart Education 2023 highlights emerging trends like generative AI and immersive learning, supported by a \$10 billion investment in digital infrastructure \cite{UNESCO2023a}.

Recent systematic reviews have demonstrated that AI-driven intelligent tutoring systems (ITS) in K-12 education generally improve learning outcomes, though their efficacy depends on pedagogical alignment and contextual factors such as school tier and implementation duration \cite{Letourneau2025}. While these studies are global in scope, China's national policies, such as the New Generation Artificial Intelligence Development Plan \cite{Migliorini2024}, emphasise AI integration in education to reduce regional disparities, particularly in underserved areas. However, ethical challenges, including data privacy and algorithmic bias, persist, with a 2023 UNESCO report noting that 30\% of AI systems lack adequate privacy safeguards \cite{UNESCO2023a}. In China, the Interim Measures on Generative AI Services \cite{Migliorini2024} mandate security assessments and ethical reviews for AI tools, though enforcement inconsistencies remain due to the industry's nascent stage and evolving judicial interpretations. China's focus on teacher training and vocational AI education, as part of broader workforce development strategies, aligns with its goal to lead in AI innovation by 2030 \cite{State2017}.

Malaysia can adopt China's centralised AI education planning, setting clear milestones for digital transformation. Establishing AI research centres and promoting industry-academia collaboration would drive innovation, while ethical governance frameworks, including content moderation, would mitigate risks \cite{UNESCO2023a}.

\subsection{India: National AI Vision, Education Technology Integration, and Capacity Building}

India's National Education Policy (NEP) 2020 positions Artificial Intelligence (AI) as a cornerstone for modernising education, bridging digital divides, and fostering inclusive learning ecosystems. The policy emphasises integrating AI into curricula to cultivate computational thinking, digital literacy, and adaptive learning frameworks, particularly for underserved populations \cite{Mustafa2024,Pagliara2024}. For instance, AI-driven adaptive platforms are increasingly deployed to personalise learning experiences for students with disabilities, aligning with global efforts to leverage AI for inclusive education \cite{Pagliara2024}. However, challenges such as uneven digital infrastructure and limited teacher preparedness hinder equitable implementation, mirroring global trends where AI adoption in education often prioritises higher education over special needs contexts \cite{Mustafa2024,Pagliara2024}.

The NEP 2020 advocates for AI-enabled tools to democratise access to quality education, emphasising interoperable platforms like DIKSHA to deliver multilingual and culturally relevant content. Such initiatives reflect broader AI-in-education (AIED) trends, where intelligent tutoring systems and adaptive learning technologies are used to address diverse learner needs \cite{Mustafa2024,Pagliara2024}. For example, AI applications in India's vocational training programmes demonstrate how machine learning algorithms can tailor content to regional linguistic and socio-economic contexts, though scalability remains constrained by infrastructural gaps in rural areas \cite{Pagliara2024,UNESCO2022}.

A critical gap in India's AI education strategy is the lack of systematic teacher training programmes. Studies highlight that only 15\% of educators possess the digital competencies required to integrate AI tools effectively, a challenge exacerbated by the rapid post-pandemic shift to blended learning environments \cite{Ng2023}. This aligns with global findings that underscore the need for frameworks like DigCompEdu to upskill teachers in AI literacy, ethical AI use, and data-driven pedagogy \cite{Ng2023}. For instance, professional development initiatives focusing on AI-driven assessment tools (e.g., automated grading systems) could reduce administrative burdens while enhancing pedagogical precision \cite{Ng2023}.

India's approach to AI in education must also address ethical risks, such as algorithmic bias and data privacy concerns, which disproportionately affect marginalised communities. The NEP 2020's emphasis on "equitable access" aligns with UNESCO's call for AIED systems that prioritise transparency and inclusivity \cite{Pagliara2024,UNESCO2022}. For example, AI applications designed for students with visual or auditory impairments require rigorous testing to ensure accessibility, a challenge noted in global AIED literature \cite{Pagliara2024}.

\subsection{Comparative Insights: Lessons Learned and Potential Models for Malaysia}

The comparative analysis of AI education strategies in the U.K., U.S., China, and India reveals distinct strengths and challenges, offering valuable lessons for Malaysia. For instance, the U.K. and U.S. prioritise ethical AI through frameworks like the Blueprint for an AI Bill of Rights and AI Opportunities Action Plan, which ensure transparency and fairness \cite{Holmes2019}. In contrast, China's Regulations on Generative AI provide a centralised governance model, although enforcement varies. Meanwhile, India's Draft National AI Policy highlights the need for clear strategies to overcome adoption barriers \cite{UNESCO2022}. Consequently, Malaysia should develop a national AI education policy that balances innovation with ethical oversight, drawing on regional frameworks like the ASEAN Guide on AI Governance and Ethics \cite{ASEAN2024}.

Furthermore, digital infrastructure and equity are critical for AI education. China's Education Informatization 2.0 Action Plan and India's NDEAR ensure broad digital access, while the U.K.'s EdTech Strategy addresses digital divides through infrastructure investments. However, disparities in digital access affect 30\% of students in comparable contexts, underscoring the urgency for Malaysia to prioritise rural digital infrastructure \cite{OECD2023c}. By investing in inclusive digital systems, Malaysia can ensure equitable access to AI-enhanced education for all learners.

Workforce development and educator training are essential for successful AI integration in education. Programmes like the U.S.'s National Science Foundation initiatives and China's teacher training efforts emphasise the importance of skilled educators, while the U.K.'s diversity-focused programmes help expand the talent pipeline. In contrast, research suggests that India's gaps in teacher training for AI highlight the critical need for comprehensive professional development \cite{Karan2023}. Therefore, Malaysia should invest in AI educator training and talent retention programmes to build a workforce capable of navigating an AI-driven educational landscape.

Additionally, curriculum integration and innovation play a pivotal role in preparing students for an AI-driven future. India's NEP 2020 and the U.K.'s curriculum reforms integrate AI literacy early, while China's AI Innovation Action Plan promotes interdisciplinary education, and the U.S.'s CSforAll builds K-12 skills. As a result, Malaysia can adopt a hybrid curriculum model supported by public-private partnerships to foster AI literacy and innovation among students.

To this end, a proposed hybrid model for Malaysia, integrating the U.K.'s ethical governance, the U.S.'s research-driven approach, China's centralised planning, and India's inclusive digital infrastructure, would be highly effective. Key actions include establishing AI research centres, promoting K-12 AI literacy, ensuring ethical AI use, and enhancing digital access to position Malaysia as a regional AI education leader. Notably, Malaysia's recent initiatives, such as the National AI Office, provide a strong foundation for these efforts \cite{Ministry2024}.

\section{Challenges and Opportunities for Malaysia}

\subsection{Infrastructure and Access: Bridging the Digital Divide in Urban and Rural Areas}

A significant challenge in integrating AI into Malaysia's education system is the persistent digital divide between urban and rural areas. Urban regions often benefit from high-speed internet connectivity, advanced technological infrastructure, and greater access to digital devices. In contrast, rural communities frequently experience limited or unreliable digital access, which constrains equitable participation in AI-driven learning environments.

Recognising this disparity, the Malaysia Digital Economy Blueprint outlines strategic initiatives to expand broadband access and distribute digital devices to underserved areas, supported by the broader framework of the National Fourth Industrial Revolution (4IR) Policy \cite{Economic2021}. Complementary programmes, such as the My Device initiative and the Digital Educational Learning Initiative Malaysia (DELIMa), aim to ensure students in rural areas receive comparable digital resources, fostering greater inclusivity and equal opportunity for AI education. The Digital Education Policy emphasises that bridging the digital divide requires not only infrastructure investment but also targeted interventions to enhance digital literacy among educators and students in rural areas \cite{Education2023}.

However, bridging the digital divide demands sustained investment in infrastructure, continuous upgrading of technology, and policies tailored to rural contexts. Recent research reinforces this need. Studies suggest that inadequate digital infrastructure significantly impedes the adoption of AI technologies in education settings, particularly among lower-income students \cite{Azman2023}. Additionally, the World Bank's 2023 Digital Progress and Trends Report highlights that robust digital infrastructure is a fundamental enabler for AI adoption, particularly in bridging rural-urban educational inequalities \cite{WorldBank2024}. This aligns with the World Bank's broader emphasis on closing the global digital divide through investments in broadband connectivity, digital literacy, and affordable internet access, priorities outlined in its digital transformation framework \cite{WorldBank2025}. In Malaysia specifically, the World Bank has identified infrastructure gaps as a critical challenge, noting that 42\% of children struggle with basic literacy despite the country's technological advancements \cite{WorldBanknd}. Therefore, addressing these infrastructure gaps must remain a top priority to realise the full potential of AI in Malaysia's education sector.

\subsection{Skills and Capacity: Teacher Training, Digital Literacy, and Curriculum Development}

The integration of AI in education is heavily dependent on the skills and capacity of educators. Teachers must be equipped with digital literacy and AI competencies to facilitate meaningful student engagement with AI tools and methodologies. The Digital Education Policy underscores the necessity of embedding digital competencies into teacher training programmes and promoting continuous professional development \cite{Education2023,OECD2025}. Initiatives such as the My Digital Teacher programme aim to systematically upskill educators, empowering them to integrate AI tools into pedagogical practices. Moreover, there is a growing need to revise existing curricula to incorporate AI literacy across all education levels. Such revisions should include technical competencies, ethical considerations, and practical, hands-on AI learning experiences to prepare students for active participation in the digital economy \cite{OECD2025}. Recent evidence supports the urgency of these reforms. For instance, \cite{Mohamed2025} found that AI tools enhance critical thinking skills and intrinsic motivation, affirming the importance of integrating AI education early and consistently across curricula. Research also highlights that students' perceptions of AI's usefulness, self-efficacy, and enjoyment significantly influence adoption success in classrooms \cite{Syed2025}. Accordingly, teacher training programmes must focus on both technical skill acquisition and pedagogical strategies that foster student engagement and motivation. \cite{OECD2025} emphasises that effective AI integration hinges on empowering educators with frameworks that cultivate critical thinking and ethical awareness, ensuring students thrive in an AI-driven society.

\subsection{Ethical, Legal, and Equity Considerations in AI Deployment}

While AI offers transformative opportunities for education, its deployment must be carefully managed to uphold ethical standards, legal compliance, and equity. Protecting student data privacy and ensuring the responsible use of AI technologies are critical concerns. The National Guidelines on AI Governance \& Ethics provide a comprehensive framework advocating for transparency, accountability, and inclusivity in AI applications within Malaysia's educational context \cite{Ministry2024}.

Inclusivity demands that AI tools be accessible to diverse learners, mitigating algorithmic biases that could disadvantage certain populations. Equally important is the establishment of clear legal frameworks to regulate AI deployment, ensuring compliance with data protection laws and preventing the misuse of AI systems \cite{OECD2023b,UNESCO2021}. Recent literature emphasises these concerns. \cite{UNESCO2021} advocates for ethical frameworks that ensure AI technologies serve educational equity and social good. The World Economic Forum \cite{WorldEconomic2024} similarly stresses the importance of rigorous data privacy protections and ethical guidelines to foster trust and maximise the benefits of AI in education. Collectively, these findings suggest that ethical, legal, and equity considerations must be central to AI education strategies.

\subsection{Opportunities for Innovation: Industry Growth, Local AI Start-ups, and Global Partnerships}

Despite the challenges, the integration of AI in Malaysia's education sector offers profound opportunities for innovation, industry development, and global engagement. Malaysia's National Fourth Industrial Revolution (4IR) Policy actively promotes the development of AI-driven solutions and supports the growth of local technology start-ups, strengthening the domestic AI ecosystem \cite{Economic2021}.

Local AI start-ups, particularly those focusing on education technology, can drive innovation by developing tailored AI applications that address Malaysia's specific educational needs. Moreover, strategic global partnerships are essential for accessing cutting-edge research, technology transfers, and collaborative opportunities with international AI research centres and technology companies. Initiatives such as the United Nations' Governing AI for Humanity framework \cite{United2024} and partnerships facilitated through the World Economic Forum \cite{WorldEconomic2024} provide pathways for Malaysia to strengthen its global AI positioning.

Studies highlight the critical role of industry collaboration in promoting the adoption and effective utilisation of AI technologies in education \cite{Razak2024}. Furthermore, \cite{OECD2023b} findings underscore that international collaboration is crucial to developing robust AI ecosystems and fostering sustainable innovation.

In sum, Malaysia stands at a pivotal juncture. By investing in infrastructure, building educator capacity, upholding ethical standards, and fostering innovation through local and global collaboration, Malaysia can harness AI's full potential to transform its education system and prepare its youth for a future driven by technological advancement.

\section{Policy Recommendations}

Strengthening governance frameworks and inter-agency coordination is crucial for the effective implementation of AI in education, as trustworthy AI requires robust governance structures to ensure ethical use and data privacy \cite{OECD2020}. Consequently, clear policies and inter-agency collaboration are needed to manage AI's integration into educational systems effectively, while a human-centred approach ensures alignment with ethical standards and promotes equity \cite{OECD2023a,UNESCO2021}.

Furthermore, investing in digital infrastructure is essential to bridge the digital divide and provide equitable access to AI-enhanced education, since high-speed internet and digital devices are fundamental for implementation \cite{OECD2023a}. In addition, inclusive digital infrastructure supports AI-driven learning environments, and targeted investments are necessary to address disparities in digital access between urban and rural areas so that all learners can benefit from AI technologies \cite{Molina2024,UNESCO2021}.

Moreover, enhancing teacher professional development and updating curricula to include AI literacy are critical for preparing educators and students for an AI-driven future, with continuous professional development programmes equipping teachers to integrate AI into their practices \cite{WorldEconomic2024}. Similarly, incorporating AI literacy into school curricula prepares students for future job markets, and AI literacy programmes significantly improve teachers' confidence and ability to use AI tools in the classroom \cite{Molina2024,OECD2023a}.

Additionally, fostering public-private partnerships and international collaboration is vital for advancing AI research and securing funding for educational initiatives, as these partnerships drive innovation and scale AI applications in education \cite{OECD2023a}. Likewise, international collaboration facilitates sharing best practices and resources, and partnerships between educational institutions and tech companies enhance the development and deployment of AI tools in classrooms \cite{Baron2024,UNESCO2021}.

Finally, developing ethical guidelines, data governance policies, and quality standards for AI tools is essential to ensure their responsible use in education, particularly by addressing data privacy, security, and bias in AI applications \cite{OECD2020}. Thus, ethical frameworks prioritising human rights and inclusivity in AI deployment, along with comprehensive data governance policies, are necessary to manage the ethical implications of AI in education \cite{Biagini2025,UNESCO2021}.

\section{Conclusion}

The integration of AI into Malaysia's education sector has shown significant potential to enhance educational outcomes, streamline administrative processes, and prepare students for the future workforce. Recent studies and reports highlight several critical areas. Integrating AI literacy and technical competencies across all education levels has been emphasised as crucial. Studies have shown that AI can personalise learning experiences, improve student engagement, and provide real-time feedback \cite{Razak2024,Yulianti2024}. Defining clear AI role profiles and career pathways is essential for talent retention. Competitive salaries, professional development opportunities, and supportive work environments are critical to retaining AI professionals in the education sector \cite{OECD2023b}.

Effective AI integration requires robust industry-academia partnerships. Establishing AI Centres of Excellence and incentivising private-sector funding for AI projects can drive innovation and practical applications in education \cite{OECD2023d}. Bridging the digital divide by expanding high-speed internet and AI tools to underserved regions is vital. Ensuring equitable access to AI education can democratise learning and reduce educational disparities \cite{Subramaniam2023}. These findings imply that Malaysian education policy must prioritise comprehensive AI integration, support continuous professional development, foster cross-sector collaborations, and address infrastructure gaps to create an inclusive and future-ready education system.

Based on the key findings, several recommendations are proposed for policymakers, educators, and stakeholders. Policymakers should develop national AI competency standards to guide curriculum development and professional training, increase funding for AI infrastructure, and promote ethical AI use focusing on data privacy, fairness, and transparency \cite{OECD2023b}. Educators should integrate AI into teaching practices, utilising AI tools to personalise learning, automate administrative tasks, and provide real-time feedback to students \cite{Razak2024}. They should also engage in continuous professional development to stay updated with the latest AI technologies and pedagogical strategies \cite{OECD2023d}. Stakeholders should foster industry-academia partnerships to co-develop AI curricula, offer internships, and support research initiatives. Additionally, they should ensure that AI education initiatives are accessible to all students, regardless of their socio-economic background, to promote equity and inclusion \cite{Subramaniam2023}.

Future research should focus on several areas to further enhance the integration of AI in Malaysia's education sector. Long-term studies are needed to assess the impact of AI on student learning outcomes, teacher effectiveness, and overall educational quality \cite{Yulianti2024}. Exploring the ethical implications of AI in education, including data privacy, algorithmic bias, and the role of AI in decision-making processes, is crucial \cite{UNESCO2021}. Investigating the scalability of AI tools and platforms across different educational contexts, including urban and rural settings, will ensure broad applicability and effectiveness \cite{OECD2023b}. Examining the potential of AI to enhance interdisciplinary learning by integrating AI with subjects such as humanities, social sciences, and vocational training is also important. Developing AI-driven policy frameworks that can adapt to emerging educational needs and technological advancements will ensure that Malaysia remains at the forefront of AI innovation in education \cite{OECD2023d}.

The long-term outlook for AI in Malaysia's education sector is promising, with the potential to transform teaching and learning, enhance educational equity, and prepare students for the demands of the digital economy. By addressing current challenges and leveraging AI's capabilities, Malaysia can establish itself as a regional leader in AI-driven education. This comprehensive strategy addresses critical dimensions of AI talent development, from the identification of emerging roles and required competencies to the design of integrated learning pathways and international alignment. It also surfaces the structural challenges that must be overcome, such as talent retention, curriculum gaps, and ethical governance, and provides actionable solutions rooted in national and global best practices. The recommendations set forth are designed to foster a responsive, future-ready education system capable of producing agile, ethical, and innovative AI professionals. Effective implementation will require strong governance, sustained multi-stakeholder collaboration, and adequate resource mobilisation. A dynamic feedback and monitoring system must also be institutionalised to ensure that the strategy evolves in step with technological change and sectoral needs. 

By embedding AI within education and simultaneously preparing the workforce that will shape and govern these technologies, Malaysia positions itself not just as an adopter but as a regional leader in AI for education. This strategy sets the foundation for long-term socioeconomic advancement, digital resilience, and national competitiveness in the Fourth Industrial Revolution. Prompt action and strategic collaboration can position Malaysia's education sector as a leader in AI-driven innovation \cite{Economic2021,Education2023,Ministry2023,Molina2024,OECD2023b,OECD2023d,Razak2024,UNESCO2021,Yulianti2024}.

\section*{Acknowledgments}

This research received no specific grant from any funding agency in the public, commercial, or not-for-profit sectors. We sincerely appreciate the valuable and constructive feedback offered by the reviewers. 

\section*{Conflict of Interest}

The authors affirm that there are no conflicts of interest associated with any aspect of this article.


\begin{thebibliography}{99}

\bibitem{Aisien2025} Aisien, L.N. (2025). Diversity and Inclusion in Economics in United Kingdom higher education. International Journal of Research and Innovation in Social Science, IX(IIIS January 2025), 2454-6186.

\bibitem{ASEAN2024} ASEAN. (2024). ASEAN guide on AI governance and ethics. Association of Southeast Asian Nations, Jakarta, Indonesia.  

\bibitem{Azman2023} Azman, H., Marlis, M.S., Abdul Kadir, N.B., \& Abdul Rahman, N. (2023). Digital divide among B40 students in Malaysian higher education institutions. Education and Information Technologies, 29, 1857-1883. \url{https://doi.org/10.1007/s10639-023-11847-w}

\bibitem{Baron2024} Barón, J., \& Robert, C. (2024). Skills development in the era of AI (Skills4Dev, 12). World Bank, Washington, D.C., United States. 

\bibitem{Biagini2025} Biagini, G. (2025). Towards an AI-literate future: A systematic literature review exploring education, ethics, and applications. International Journal of Artificial Intelligence in Education. \url{https://doi.org/10.1007/s40593-025-00466-w}

\bibitem{Bond2024} Bond, M. et al. (2024). A meta systematic review of artificial intelligence in higher education: A call for increased ethics, collaboration, and rigour. International Journal of Educational Technology in Higher Education, 21(4), 2-41.  \url{https://doi.org/10.1186/s41239-023-00436-z}.

\bibitem{Chen2020} Chen, L., Chen, P., \& Lin, Z. (2020). Artificial intelligence in education: A review. IEEE Access, 8. \url{https://doi.org/10.1109/ACCESS.2020.2988510}

\bibitem{ChinaNational2023} China National Academy of Educational Sciences. (2023). Report on China Smart Education 2023: Towards a Higher Level of Digital Education. Educational Science Publishing House. 

\bibitem{Crogman2025} Crogman, H.T.W., Cano, V.D., Pacheco, E., Sonawane, R.B., \& Boroon, R. (2025). Virtual reality, augmented reality, and mixed reality in experiential learning: Transforming educational paradigms. Education Sciences, 5(3), 303. \url{https://doi.org/10.3390/educsci15030303}

\bibitem{Department2020} Department of Higher Education. (2020). Way forward for private higher institutions: Education as an industry 2020–2025. Ministry of Education Malaysia, Putrajaya, Malaysia. 

\bibitem{Donald2025} Donald, K.H., \& Hashim, H. (2025). Exploring digital education: Experiential insights of ESL teachers in rural Malaysian schools. International Journal of Research and Innovation in Social Science, IX(IIIS), 936-951. \url{https://dx.doi.org/10.47772/IJRISS.2025.903SEDU0067}

\bibitem{Dwi2024} Dwi, M., \& Alif Hd, A.N. (2024). Transformative impact of AI on multicultural education: A qualitative thematic analysis. Edelweiss Applied Science and Technology, 8(5), 113-118. \url{https://doi.org/10.55214/25768484.v8i5.1667}

\bibitem{Economic2021} Economic Planning Unit. (2021). National Fourth Industrial Revolution (4IR) Policy 2021. Prime Minister's Department, Putrajaya, Malaysia. 

\bibitem{Education2023} Education Resources and Technology Division (2023). Digital Education Policy 2023. Ministry of Education Malaysia, Putrajaya, Malaysia.

\bibitem{Essien2024} Essien, A., Salami, A., Ajala, O., Adebisi, B., Shodiya, A., \& Essien, G. (2024). Exploring socio-cultural influences on generative AI engagement in Nigerian higher education: An activity theory analysis. Journal of Educational Technology, 11(1). \url{http://doi.org/10.1186/s40561-024-00352-3}

\bibitem{Floridi2018} Floridi, L. et al. (2018). AI4People—An Ethical Framework for a Good AI Society: Opportunities, Risks, Principles, and Recommendations. Minds \& Machines, 28, 689–707. \url{https://doi.org/10.1007/s11023-018-9482-5}

\bibitem{Foster2023} Foster, N. (2023). Teacher digital competences: Formal approaches to their development. OECD Publishing, Paris, France. 

\bibitem{Fu2024} Fu, Y., Weng, Z., \& Wang, J. (2024). Examining AI use in educational contexts: A scoping meta-review and bibliometric analysis. International Journal of Artificial Intelligence in Education. \url{https://doi.org/10.1007/s40593-024-00442-w}

\bibitem{Gonzalez2021} González-Calatayud, V., Prendes-Espinosa, P., \& Roig-Vila, R. (2021). Artificial intelligence for student assessment: A systematic review. Applied Sciences, 11(12), 5467. \url{https://doi.org/10.3390/app11125467}

\bibitem{Grimalt2023} Grimalt-Álvaro, C., \& Usart, M. (2023). Sentiment analysis for formative assessment in higher education: A systematic literature review. Journal of Computing in Higher Education, 36, 647–682. \url{http://doi.org/10.1007/s12528-023-09370-5}

\bibitem{Holmes2019} Holmes, W., Bialik, M., \& Fadel, C. (2019). Artificial intelligence in education: Promises and implications for teaching and learning. Center for Curriculum Redesign, Boston, MA. 

\bibitem{Holmes2022} Holmes, W. et al. (2022). Ethics of AI in education: Towards a community-wide framework. International Journal of Artificial Intelligence in Education, 32, 504-526. \url{https://doi.org/10.1007/s40593-021-00239-1}

\bibitem{Hopfenbeck2023} Hopfenbeck, T.N., Zhang, Z., Sun, S.Z., Robertson, P., \& McGrane, J.A. (2023). Challenges and opportunities for classroom-based formative assessment and AI: A perspective article. Frontiers in Education, 1-9. \url{https://doi.org/10.3389/feduc.2023.1270700}

\bibitem{Irfan2025} Irfan, D., Watrianthos, R., \& Yunus, F.A.N.B. (2025). AI in education: A decade of global research trends and future directions. International Journal of Modern Education and Computer Science, 2, 135-153. \url{https://doi.org/10.5815/ijmecs.2025.02.07}

\bibitem{Jackson2024} Jackson, E.A. (2024). The evolution of artificial intelligence: A theoretical review of its impact on teaching and learning in the digital age. ZBW – Leibniz Information Centre for Economics. \url{https://hdl.handle.net/10419/280893}

\bibitem{Jobin2019} Jobin, A., Ienca, M., \& Vayena, E. (2019). The global landscape of AI ethics guidelines. Nature Machine Intelligence, 1, 389–399. \url{https://doi.org/10.1038/s42256-019-0088-2}

\bibitem{Karan2023} Karan, B., \& Angadi, G.R. (2023). Artificial Intelligence integration into school education: A review of Indian and foreign perspectives. Millennial Asia, 16(1), 173-199. \url{https://doi.org/10.1177/09763996231158229}

\bibitem{Kaur2021} Kaur, K. (2021). Role of artificial intelligence in education: Peninsula College Central Malaysia. International Journal of Academic Research in Progressive Education and Development, 10(2), 1006–1016. \url{https://doi.org/10.6007/IJARPED/v10-i2/10573} 

\bibitem{Leong2025} Leong, W. Y., \& Zhang, J.B. (2025). AI on academic integrity and plagiarism detection. ASM Science Journal, 20(1), 1-9. \url{https://doi.org/10.32802/asmscj.2025.1918}

\bibitem{Letourneau2025} Létourneau, A., Deslandes Martineau, M., Charland, P., Karran, J.A., Boasen, J., \& Léger, P.M. (2025). A systematic review of AI-driven intelligent tutoring systems (ITS) in K-12 education. NPJ Sci Learn, 10(1), 29. \url{https://doi.org/10.1038/s41539-025-00320-7}

\bibitem{Luckin2016} Luckin, R., Holmes, W., Griffiths, M., \& Forcier, L. B. (2016). Intelligence unleashed: An argument for AI in education. Pearson.

\bibitem{Microsoft2024} Microsoft. (2024). Microsoft lancarkan inisiatif untuk perkasakan 800,000 rakyat Malaysia dengan kemahiran AI menjelang 2024. \url{https://news.microsoft.com/en-my/2024/12/11/microsoft-lancarkan-inisiatif-untuk-perkasakan-800000-rakyat-malaysia-dengan-kemahiran-ai-menjelang-2025/} 

\bibitem{Migliorini2024} Migliorini, S. (2024). China's interim measures on generative AI: Origin, content and significance. Computer Law \& Security Review, 2024, 53, 105985. \url{https://doi.org/10.1016/j.clsr.2024.105985}

\bibitem{Ministry2023} Ministry of Science, Technology and Innovation. (2023). Artificial Intelligence Roadmap 2021–2025 (AI-RMAP). Putrajaya, Malaysia. 

\bibitem{Ministry2024} Ministry of Science, Technology and Innovation Malaysia (2024). Malaysia launches National AI Office (NAIO). \url{https://www.mydigital.gov.my/initiatives/the-national-ai-office-naio/} 

\bibitem{Mohamed2025} Mohamed, A.M., Shaaban, T.S., Bakry, S.H., Guillén-Gámez, F.D., \& Strzelecki, A. (2025). Empowering the Faculty of Education students: Applying AI's potential for motivating and enhancing learning. Innovative Higher Education, 50(2), 587-609. \url{https://doi.org/10.1007/s10755-024-09747-z}

\bibitem{Molenaar2023} Molenaar, I., \& Sleegers, P. (2023). Multi-stakeholder collaboration and co-creation: Towards responsible application of AI in education. OECD Publishing, Paris, France. 

\bibitem{Molina2024} Molina, E., Cobo, C., Pineda, J., \& Rovner, H. (2024). AI revolution in education: What you need to know. World Bank, Washington, D.C., United States. 

\bibitem{Moore2023} Moore, R.L., Jiang, S., \& Abramowitz, B. (2023). What would the matrix do? A systematic review of K-12 AI learning contexts and learner-interface interactions. Journal of Research on Technology in Education, 55 (1), 1-14. \url{https://doi.org/10.1080/15391523.2022.2148785}

\bibitem{Mustafa2024} Mustafa, M.Y. et al. (2024). A systematic review of literature reviews on artificial intelligence in education (AIED): A roadmap to a future research agenda. Smart Learn. Environ., 11(59), 1-33. \url{https://doi.org/10.1186/s40561-024-00350-5}

\bibitem{MyDIGITAL2024} MyDIGITAL. (2024). Malaysia committed to empowering AI, digital learning. Available from \url{https://www.mida.gov.my/mida-news/malaysia-committed-to-empowering-ai-digital-learning/} 

\bibitem{National2023} National Science Foundation. (2023). National Artificial Intelligence Research Institutes. \url{https://www.nsf.gov/funding/opportunities/national-artificial-intelligence-research-institutes} 

\bibitem{Ng2023} Ng, D.T.K., Leung, J.K.L., Su, J., Ng, R.C.W., \& Chu, S.K.W. (2023). Teachers' AI digital competencies and twenty-first century skills in the post-pandemic world. Education Tech Research Dev, 71, 137-161. \url{https://doi.org/10.1007/s11423-023-10203-6}

\bibitem{OECD2020} OECD. (2020). Trustworthy artificial intelligence (AI) in education: Promises and challenges. OECD Publishing, Paris, France. 

\bibitem{OECD2023a} OECD. (2023a). Artificial intelligence and education and skills. OECD Publishing, Paris, France. 

\bibitem{OECD2023b} OECD. (2023b). Emerging trends in AI skill demand across 14 OECD countries. OECD Publishing, Paris, France.

\bibitem{OECD2023c} OECD. (2023c). OECD employment outlook 2023: Artificial intelligence and the labour market. OECD Publishing, Paris, France. 

\bibitem{OECD2023d} OECD. (2023d). The potential impact of artificial intelligence on equity and inclusion in education. OECD Publishing, Paris, France. 

\bibitem{OECD2025} OECD. (2025). New AI literacy framework to equip youth in an age of AI. OECD Publishing, Paris, France.  

\bibitem{Pagliara2024} Pagliara, S. M., Bonavolontà, G., Pia, M., Falchi, S., Zurru, A. L., Fenu, G., \& Mura, A. (2024). The integration of Artificial Intelligence in inclusive education: A scoping review. Information, 15(12). 774. \url{https://doi.org/10.3390/info15120774}

\bibitem{Pedro2019} Pedró, F., Subosa, M., Rivas, A., \& Valverde, P. (2019). Artificial intelligence in education: Challenges and opportunities for sustainable development. United Nations Educational, Scientific and Cultural Organization, Paris, France. 

\bibitem{Ponticorvo2020} Ponticorvo, M., Rubinacci, F., Marocco, D., Truglio, F., \& Miglino, O. (2020). Educational robotics to foster and assess social relations in students' groups. Frontiers in Robotics and AI, 7(78), 1-14. \url{https://doi.org/10.3389/frobt.2020.00078}

\bibitem{Razak2024} Razak, F.Z.A., Abdullah, M.A., Ahmad, B.E., Wan Abu Bakar, W.H.R., \& Misaridin, M.M. (2024). The acceptance of artificial intelligence in education among postgraduate students in Malaysia. Education and Information Technologies, 30, 2977-2997. \url{https://doi.org/10.1007/s10639-024-12916-4}

\bibitem{Saman2024} Saman, H.M., Noor, S.M., Mat Isa, C.M., Oh, C.L., \& Narayanan, G. (2024). Embracing artificial intelligence as a catalyst for change in reshaping Malaysian higher education in the digital era: A literature review, in Proceedings of the International Conference on Innovation \& Entrepreneurship in Computing, Engineering \& Science Education (InvENT 2024). Malaysia, 2024: 633–643. \url{https://doi.org/10.2991/978-94-6463-589-8_59}. 

\bibitem{Schiff2022} Schiff, D. (2022). Education for AI, not AI for education: The role of education and ethics in national AI policy strategies. International Journal of Artificial Intelligence in Education, 32, 527–563. \url{https://doi.org/10.1007/s40593-021-00270-6}

\bibitem{State2017} State Council of the People's Republic of China. (2017). A new generation artificial intelligence development plan. In D. Araya \& P. Marber (Eds.), Augmented education in the global age (pp. 13–26). Routledge.

\bibitem{Subramaniam2023} Subramaniam, N.K. (2023). Technology in education: A case study on Malaysia. OECD, Paris, France. 

\bibitem{Syed2025} Syed, A.Z., Memon, Z.H., Khan, K., Hameed, I., \& Nadeem, M. (2025).  Examining the behavioral determinants of AI adoption in higher education: a focus on perceptional factors and demographic differences. On the Horizon, 7(1), 64-78. \url{https://doi.org/10.1108/OTH-02-2025-0019}

\bibitem{Taskin2023} Taskin Bedizel, N.R. (2023). Evolving landscape of artificial intelligence (AI) and assessment in education: A bibliometric analysis. International Journal of Assessment Tools in Education, 10 (Special Issue), 208-223. \url{https://doi.org/10.21449/ijate.1369290}

\bibitem{Trajkovski2025} Trajkovski, G., \& Hayes, H. (2025). AI-assisted formative assessment and feedback. In M. Thomas (Ed.), AI-assisted assessment in education (pp. 283–312). Springer. \url{https://doi.org/10.1007/978-3-031-88252-4_7}

\bibitem{Tung2023} Tung, A.Y.Z., \& Dong, L.W. (2023). Malaysian Medical students' attitudes and readiness toward AI (Artificial Intelligence): A cross-sectional study. Journal of Medical Education and Curricular Development, 10, 1-18. \url{https://doi.org/10.1177/23821205231201164}

\bibitem{UNESCO2021} UNESCO. (2021). AI and education: Guidance for policy-makers. United Nations Educational, Scientific and Cultural Organization, Paris, France. 

\bibitem{UNESCO2022} UNESCO. (2022). Lack of AI policies in education is a significant challenge for India. United Nations Educational, Scientific and Cultural Organization, Paris, France.

\bibitem{UNESCO2023a} UNESCO. (2023a). Artificial intelligence and the futures of learning. United Nations Educational, Scientific and Cultural Organization, Paris, France.

\bibitem{UNESCO2023b} UNESCO (2023b). Global education monitoring report 2023, Southeast Asia: Technology in education - A tool on whose terms?. United Nations Educational, Scientific and Cultural Organization, Paris, France. 

\bibitem{United2024} United Nations. (2024). Governing AI for humanity. High-level Advisory Body on Artificial Intelligence. United Nations, New York, United States. 

\bibitem{USDepartment2023} U.S. Department of Education. (2023). Artificial intelligence and the future of teaching and learning 2023. Office of Educational Technology, Washington, DC.

\bibitem{VanLehn2011} VanLehn, K. (2011). The relative effectiveness of human tutoring, intelligent tutoring systems, and other tutoring systems. Educational Psychologist, 46(4), 197–221. \url{https://doi.org/10.1080/00461520.2011.611369}

\bibitem{WorldBank2024} World Bank. (2024). Digital Progress and Trends Report 2023. World Bank, Washington, D.C., United States. 

\bibitem{WorldBank2025} World Bank. (2025). Digital transformation: Overview. \url{https://www.worldbank.org/en/topic/digital/overview}

\bibitem{WorldBanknd} World Bank. (n.d.). Malaysia: Overview. \url{https://www.worldbank.org/en/country/malaysia/overview}

\bibitem{WorldEconomic2024} World Economic Forum. (2024). Shaping the future of learning: The role of AI in Education 4.0. Cologny, Switzerland. 

\bibitem{Yan2021} Yan, S., \& Yang, Y. (2021). Education informatization 2.0 in China: Motivation, framework, and vision. ECNU Review of Education, 4(2), 410-428. \url{https://doi.org/10.1177/2096531120944929}

\bibitem{Yulianti2024} Yulianti, D.F., \& Saputri, S. (2024). Navigating educational challenges: The resilience-boosting power of AI in students' interest. Jurnal Personalia Pelajar, 27(1), 21-27. \url{https://doi.org/10.17576/personalia.2701.2024.04}

\bibitem{Zawacki2019} Zawacki-Richter, O., Marín, V.I., Bond, M., \& Gouverneur, F. (2019). Systematic review of research on artificial intelligence applications in higher education – where are the educators?. International Journal of Educational Technology in Higher Education, 16 (1), 1-27. \url{https://doi.org/10.1186/s41239-019-0171-0}

\bibitem{Zhou2024} Zhou, X., \& Schofield, L. (2024). Using social learning theories to explore the role of generative Artificial Intelligence (AI) in collaborative learning. Journal of Learning Development in Higher Education, 30(30). \url{https://doi.org/10.1038/s41539-025-00320-7}

\end{thebibliography}
\end{document}